\documentclass[a4paper,fleqn]{cas-dc}
\usepackage{multirow}
\usepackage[numbers]{natbib}
\usepackage{subfigure}
\usepackage[center]{caption}
\usepackage[switch]{lineno}
 
\usepackage{xcolor}

\def\tsc#1{\csdef{#1}{\textsc{\lowercase{#1}}\xspace}}
\tsc{WGM}
\tsc{QE}
\tsc{EP}
\tsc{PMS}
\tsc{BEC}
\tsc{DE}

\begin{document}
\let\WriteBookmarks\relax
\def\floatpagepagefraction{1}
\def\textpagefraction{.001}

\shorttitle{Predictive Analysis on Cyber Security Threats}

\shortauthors{Fatama et~al.}

\title [mode = title]{A Data-Driven Predictive Analysis on Cyber Security Threats with Key Risk Factors}                      
\author[1]{Fatama Tuz Johora}






\affiliation[1]{organization={Department of Computer Science and Engineering, University of Chittagong},
    addressline={Chattogram-4331}, 
    country={Bangladesh}}

\author[2]{Md Shahedul Islam  Khan}
\affiliation[2]{organization={Department of School of Electronics and Information, Northwestern Polytechnical University},
    city={Xi'an, Shaanxi},
    country={China}}

\author[1]{Esrath Kanon}


\author[3]{Mohammad Abu Tareq Rony}
\affiliation[3]{organization={Department of Statistics, Noakhali Science and Technology University},
    addressline={Noakhali-3814}, 
    country={Bangladesh}}

\author[4]{Md Zubair}[orcid=0000-0001-7384-2805]
\cormark[1]
\ead{zubairhossain773@gmail.com}
\affiliation[4]{organization={Department of Computer Science and Engineering, Chittagong University of Engineering \& Technology},
    addressline={Chattogram-4349}, 
    country={Bangladesh}}

\author[5]{Iqbal H. Sarker}[orcid=0000-0003-1740-5517]
\cormark[1]
\ead{m.sarker@ecu.edu.au}
\affiliation[5]{organization={Centre for Securing Digital Futures, Edith Cowan University},
    addressline={Perth, WA-6027}, 
    country={Australia}}

\cortext[cor1]{Corresponding author}



\begin{abstract}
Cyber risk refers to the risk of defacing reputation, monetary losses, or disruption of an organization or individuals, and this situation usually occurs by the unconscious use of cyber systems. The cyber risk is unhurriedly increasing day by day and it is right now a global threat. Developing countries like Bangladesh face major cyber risk challenges. The growing cyber threat worldwide focuses on the need for effective modeling to predict and manage the associated risk. This paper exhibits a Machine Learning(ML) based model for predicting individuals who may be victims of cyber attacks by analyzing socioeconomic factors. We collected the dataset from victims and non-victims of cyberattacks based on socio-demographic features. The study involved the development of a questionnaire to gather data, which was then used to measure the significance of features. Through data augmentation, the dataset was expanded to encompass 3286 entries, setting the stage for our investigation and modeling. Among several ML models with 19, 20, 21, and 26 features, we proposed a novel Pertinent Features Random Forest (RF) model, which achieved maximum accuracy with 20 features (95.95\%) and also demonstrated the association among the selected features using the Apriori algorithm with Confidence (above 80\%) according to the victim. We generated 10 important association rules and presented the framework that is rigorously evaluated on real-world datasets, demonstrating its potential to predict cyberattacks and associated risk factors effectively. Looking ahead, future efforts will be directed toward refining the predictive model's precision and delving into additional risk factors, to fortify the proposed framework's efficacy in navigating the complex terrain of cybersecurity threats.
\end{abstract}



\maketitle

\section{Introduction}
The escalating complexity of cybersecurity threats necessitates a comprehensive understanding of their underlying dynamics. The use of technology has rapidly transformed the world into a more efficient and mobile place. However, with the increasing reliance on technology, the threat of cyber-attacks has become a significant concern for organizations and individuals alike. The Symantec Annual Threat Report of 2017 revealed a 10\% increase in targeted attacks on organizations, indicating the harmful usage of the internet and social media for malicious purposes \cite{R1}. Cyber risk typically refers to any risk connected to economic loss, business interruption, or deface to the prestige of an organization or individual owing to unauthorized, inappropriate, or inaccurate use of information systems. The increasing trend of cyber-attacks has increased spear-phishing emails (71\%) and new malware variants (87.7\%) as reported in 2018 \cite{R2}. The critical infrastructure has become a prime target for cyber attackers, causing not only inconvenience but also life-threatening situations. According to the Thales Global Security report of 2018, 36\% of significant enterprises worldwide experienced a data contravention, with only the United States accounting for 46\% of these breaches \cite{R3}. The cost of insufficient cyber security is estimated to have reached USD 945 billion globally in 2020 \cite{R4}.

Despite the growing importance of cyber security, data on cyber threats remains scarce. The Threat Report of 2024 revealed that for Data Breaches, 28\% experiencing a Ransomware attack \cite{R5}. Human error, such as falling victim to phishing campaigns or accessing harmful websites, accounted for exposing over two billion records, by the IBM Threat Intelligence Index Report of 2018 \cite{R6}. The issue of cyber attacks is particularly concerning in Bangladesh, where many individuals are not aware of the dangers posed by such threats. Careless behaviour often leads to victims falling into cyber risk, often leading to devastating consequences such as suicide. The World Economic Forum's Global Risk Report mentioned critical infrastructure cyber attacks as the fifth greatest risk \cite{R7}.

Data-driven technologies have become a popular trending topic in computing days and the demand for it is climbing gradually. Since the majority of things in our everyday life have now been digitally captured as data and the acumen that can be gleaned from the data is the key to developing rational solutions \cite{sarker2024ai}. These data-driven solutions may be applied in a range of real-world contexts, such as substance abuse, IoT applications, financial or business analysis, and others, to build an effective model and make wise judgments \cite{shi2011optimization, sarker2024ai, olson2007introduction}.  
In this research paper, we aim to identify and analyze the cyber-based security threats faced by Bangladeshi netizens. To achieve this goal, we collected data from the target population using a generated questionnaire to create a dataset. Pre-processing was done on the dataset and the contribution of the defining characteristics to the target variable was determined. The insignificant features were removed, and the important features were utilized to develop a model for prediction based on ML classification algorithms \cite{sarker2023machine}. The attributes used in the classifier model were then explored using association rule mining to recognize key risk factors of cyber risk \cite{sarker2023multi}. Our suggested framework includes both a predictive classifier and an association rule-based technique, which is an important technology in the field of artificial intelligence \cite{sarker2023multi}. The implementation of this well-designed framework, along with identifying a significant risk factor of cyber risk through association rule mining and an in-depth study of experimental results, represents a significant contribution to the field of cyber security. The following list summed up the study's contributions:
\begin{itemize}
\item[--] To develop the questionnaires and gather information accordingly to extract real-life insights through data-driven analysis. 

\item[--] To predict socioeconomic factor-based cyber threats employing ML methods. 

\item[--] To derive a set of rules with key risk factors that result from the breakdown of notable features.

\item[--] To complete experimental analysis on actual datasets to demonstrate the model performance.
\end{itemize}

This work has been organized in the following flow, section \textcolor{red}{\ref{sec:2}} reviews related works regarding this study. Section \textcolor{red}{\ref{sec:3}}, describe whole dispenses a thorough explanation of the dataset and outlines the workflow by dividing it into different modules, section \ref{sec:4} offers an in-depth review of the performance and demonstrates the study's effectiveness, and shows the experimental results. Section \ref{sec:5} highlights some insights of the study and discussion. The final part of this paper is Section \ref{sec:6}.

\section{Background study}\label{sec:2}
Innumerable studies have explored various aspects of cyber risk, including its causes, severity, and impact on daily activities. Despite the extensive research on this topic, there remains a scarcity of research on cyber risk prevention. Furthermore, there is a dearth of research focusing on intervention strategies based on ML, with only a few relevant works available in the literature.

Amgad Muneer et al.\cite{muneer2020comparative} presented a comparative analysis of ML methods for detecting cyberbullying based on Twitter tweets.
They also evaluated the effectiveness of each classifier using performance
metrics. Frequency-Inverse Document Term Frequency (TF-IDF) and Word2Vec
were used to extricate features. Zaheer Abbass et al.\cite{abbass2020framework} developed a system for predicting
social media crime (cyberstalking, cyberbullying, cyber hacking, cyber
harassment, and cyber scam) based on Twitter tweets using ML techniques. ML classification algorithms:
Multinomial Nave Bayes (MNB), K-Nearest Neighbors (KNN), and SVM were employed to categorize several types of crime.
John Hani et al.\cite{mounir2019social} used 2 ML algorithms to detect
cyberbullying on social media platforms, such as SVM and Neural Network. TF-IDF technique was employed for feature extraction. Arun Kulkarni et al.\cite{jain2019phishing} used a machine-learning approach to detect phishing websites from 1353 samples. The primary target was to develop a system that categorizes websites according to their URLs. Author Subroto et al.\cite{subroto2019cyber} proposed a model that uses social media big data analytics and statistical ML to forecast cyber risk. Based on social media chats, the predictions were created by examining the software's susceptibilities to threats. John O. Awoyemi et al. \cite{awoyemi2017credit} presented a comparison of ML algorithms for credit card fraud detection. Brij B. Gupta et al.\cite{gupta2021novel} introduced a novel method to detect phishing URLs in a real-time environment using lexical-based ML. This method builds an ML-based phishing detection system that enables users to rapidly identify the authenticity and harm of a website.

Sarker et al. \cite{sarker2024ai} explored various classification algorithms based on ML to detect and classify cyber intrusions. ML techniques such as the Bayesian approach, the Tree-based model, and the Artificial Neural Network (ANN) based model were used to identify several types of cyber-attacks and conduct comparative tests. ML classification algorithms: Naïve Bayes (NB), KNN, and LR were used to scrutinize highly skewed credit card fraud data. Principle component analysis (PCA) was applied for feature selection. The study\cite{le2024search} assesses the impact of Search Engine Optimization (SEO) poisoning on Small and Medium-sized Enterprises (SMEs) and their digital marketing. It outlines the risks, including reputational and financial loss, and operational disruptions. Adapting the National Institute of Standards and Technology (NIST) Cybersecurity Framework, it suggests mitigation strategies for SMEs, such as security audits and training. The research provides insights into the complexities of SEO poisoning, advocating for proactive defenses and ongoing cybersecurity vigilance to protect SMEs in the digital marketing sphere.

Another group of researchers \cite{fatoki2024optimism} examines the relationship between personal optimism bias and cybersecurity behavior in organizations, revealing how this bias leads to negative cybersecurity attitudes and riskier behaviors. It also explores how information security awareness can mitigate these effects. Using a survey of non-IT U.S. employees and structural equation modeling, findings show that optimism bias undermines cybersecurity, while awareness helps correct attitudes and reduce risky actions. The research contributes to our understanding of optimism bias's role in cybersecurity and underscores the need for targeted education and training to improve non-IT staff cybersecurity practices.

\subsection{Research gap}
In the field of Cyber Security
Threats with Key Risk Factors, the literature review identifies a significant research gap in the realm of Cyber Security
Threat analysis, particularly in scalability, adaptability, computational efficiency, and real-time data. In contrast to the aforementioned works, we intend to include individuals from diverse backgrounds to create an inclusive novel dataset, thereby laying the groundwork for a comprehensive framework consisting of a predictive classifier and the identification of key risk factors via association rule mining.

\section{Methodology} \label{sec:3}
A comprehensive explanation of the study techniques and their associated workflow is provided in this section. We analyzed the novel dataset which was further used in constructing the applied methods through evaluation, employing a variety of hyperparameters to assess their performance. To offer a thorough understanding of the Cybersecurity threats, a description of the suggested approach's architecture is also included. Figure\ref{flowfig} illustrates the analytical approach proposed for this study.

\subsection{Questionnaire and data characteristics}  \label{subsec:3.1}
We created a question paper after evaluating the literature and various aspects of cybercrime and cyberthreats based on the key variables linked to cyber risk, before approving a total of 26 multiple-choice questions (MCQ). Each problem has at least two distinctive solutions. The following are the conclusive questions:
\begin{itemize}
    \item[] \textit{Q1.} Has the entrant routinely manipulated passwords such as 1,2,3,4,5,6 for their local machines like laptops, mobiles, or any online account or social networking website password?
    \item[] \textit{Q2.} Does the user frequently employ all of their social media accounts, in particular Facebook, Instagram, and Twitter?
    \item[] \textit{Q3.} Does the user disclose their sentiment on social media through status updates or private messages?
    \item[] \textit{Q4.} Has the user ever been poorly influenced by blackmailing through a fallacious online account?
    \item[] \textit{Q5.} Have they safeguarded their internet effectively?
    \item[] \textit{Q6.} Have they used several devices to access their online account?
    \item[] \textit{Q7.} Has the user shared their private information, such as images, codes, or bank account numbers, on social media or any unauthorized organization?
    \item[] \textit{Q8.} Does the user customarily receive emails from strangers or organizations?
    \item[] \textit{Q9.} Are they employing their email account password to another online account?  
    \item[] \textit{Q10.} Did the user permit ingress to their email account to anybody else?  
    \item[] \textit{Q11.} Do they habitually click on spam or unanticipated email links?
    \item[] \textit{Q12.} Does the participant regularly buy products online?
    \item[] \textit{Q13.} Has the user misplaced money by purchasing or vending commodities online?
    \item[] \textit{Q14.} Is the participant a compulsive buyer?
    \item[] \textit{Q15.} Has a partaker installed prohibited malicious software on their computer?
    \item[] \textit{Q16.} Does the user consistently share their private devices like computers, mobile phones, etc. with others?
    \item[] \textit{Q17.} Does the participant download software from authorized sources?
    \item[] \textit{Q18.} Do they access Virtual Private Networks (VPN) persistently?
    \item[] \textit{Q19.} Does the user continuously store their credentials in the browser's memory?
    \item[] \textit{Q20.} Does the participant commonly use virus-infected USB flash in their devices?
    \item[] \textit{Q21.} Does the user often keep their devices up-to-date?
    \item[] \textit{Q22.} What is the age range of the partaker?
    \item[] \textit{Q23.} What gender is the participant?
    \item[] \textit{Q24.} Did the participant grant access to their internet account to anyone else?
    \item[] \textit{Q25.} Does the participant know about cybercrimes such as phishing, stalking, bullying, malware attacks, purchase fraud, sales fraud, payment fraud, etc.?
        \item[] \textit{Q26.} Is the participant aware of cybercrime?
\end{itemize}

\begin{table*}[width=0.72\textwidth]
\caption{List of attribute names in the dataset}
\centering
\begin{tabular*}{\tblwidth}{l|l|l|l}
\hline
\hline
{\bfseries Key Factors } &  {\bfseries Question} & {\bfseries Feature Name } & {\bfseries Feature Type}\\
\hline
\hline
{\itshape Social media } & Q1 & weak-password & binary\\
{\itshape unawareness } & Q2 & social-media-user & binary\\
                         & Q3 & disclose-sentiment-on-social & \\ && -media & binary\\
                         & Q4 & victimized-by-blackmailing & binary\\
                         & Q5 & maintained-privacy-on-social & \\&& -media & binary\\
                         & Q6 & accessing-online-account  & \\
                         & & -using-several-devices  & binary\\
                         & Q7 & sharing-private-information-on  &\\ && -the-internet & binary\\
\hline
{\itshape Email user threat } & Q8 & receive-phishing-email  & binary\\
 & Q9 & shared-email-access & binary\\
                         & Q10 & permitted-ingress-in-email & binary\\
                         & Q11 & clicked-on-spam-email-links & binary\\
                         
\hline
{\itshape Purchase from online } 
                         & Q12 & online-products-purchaser & binary\\
                         & Q13 & lost-money-by-purchasing-online & \\ && -commodities & binary\\
                         & Q14 & compulsive-buyer & binary\\

\hline
{\itshape Software and Network } & Q15 & installed-malicious-software & binary\\
                         & Q16 & shared-private-devices & binary\\
                         & Q17 & {download-unauthorized-software \ } & binary\\
                         & Q18 & accessed-VPN & binary\\
                         & Q19 & stored-credentials-on-browsers & binary\\
                         & Q20 & used-virus-infected-pen-drive & binary\\
                         & Q21 & devices-keep-updated & binary\\
\hline
{\itshape Biology, Privacy  } & Q22 & age-range & discrete\\
                   {\itshape and Awareness}      & Q23 & gender & binary\\
                         & Q24 & shared-internet-account-access & binary\\
                         & Q25 & knowledge-about-cybercrime & ordinal\\
                         & Q26 & aware-about-cybercrime & binary\\

\hline
\end{tabular*}
    \label{tab1}
\end{table*}

The profound exploration of the above questionnaire: all of the questions have only two answers which are binary type questions excluding questions 22 and 25 which are discrete and ordinal types of questions represented in Table \ref{tab1}. The attributes of the dataset are finally formed by each of these questions and the responses to these questions are the main data values or records of the dataset. Data has been collected individually both from the people who have not faced cyber attacks and also those who suffered cyber attacks, and stored in two different datasets. Finally, we have combined the datasets and added a new column named "Victim" that accommodates the values "1" and "0" for the people who suffered cyber attacks and those who did not, respectively. Eventually, both individual datasets are merged to form our final dataset and this culminating dataset holds 27 features.

Characteristics of 26 features have been represented in Table \ref{tab1}. The first column (Key Factors) symbolizes the principal objective for each question,  the second column (Question) represents the sequence number of the questions, the third column indicates the feature names and the fourth column highlights the type of individual feature. Most of the significant features were identified through the literature review; moreover, some features were selected by consulting with social organizations and cyber victims. The feature column involves the response for each question.

\subsection{Data collection and pre-processing} 
Before considering the victims of cyber attacks, pertinent literature is outfitted to determine the problem's underlying habitual causes. Social media unawareness, email user danger, and the usage of hazardous software and networks were spotlighted as common factors in the research. To authenticate the outcome from affined literature, we have revised and verified the findings by cyber assaulted people presently engaged in pertinent studies. The finished causes for the creation of the questionnaire were determined after considering the diverse cyber risk factors. The questionnaire was developed to collect information on the primary cyber risk factors identified by any Bangladeshi citizen. The questionnaire incorporates 26 multiple-choice questions that cover the causes, local circumstances, and empirical results. Questions centered on the misuse of social media, the use of detrimental software and networks, the purchase of online goods, and the risks associated with email operations. We conclude our questionnaire after considering all relevant factors. The primary motive of data collecting is to determine how responses to similar questions change between victims and non-victims of cyber attacks. After collecting the replies to identical inquiries from both groups, we could determine which responses were the most divergent between the two.  
An online survey and in-person interviews were used to collect responses from victim participants and non-victim participants, respectively. The following paragraphs elaborate on the reasons behind the deployment of distinct data-gathering methods for distinct categories. Due to societal stigma and the difficulties of authentication or identification an individual as a victim, cyber-attack victims' data has been obtained through in-person interviews. Using a Google form, the information of non-victims has been gathered and transmitted to recipients through email or social media. The form comprises introductory paragraphs stating the goal of the study and requesting permission to use the participant's information for research purposes. Once responses have been received from both victims and non-victims, these responses are organized into 2 distinct spreadsheets. In the spreadsheet, each question is condensed into an attribute name that represents each column in the dataset, and the answers to every single question are replaced with numerical values between '1', '2', and '3', depending on whether the question is binary or a 3-point Likert scale question, as shown in subsection \ref{subsec:3.1}. Lastly, a variable labelled "goal" is added to both datasets to differentiate between victims and non-victims. The values within the goal column are binary: '1' and '0' are allocated to the variable 'goal' in the victim and non-victim datasets, respectively. The combination of these two datasets yields a functional data collection. This dataset has 27 variables in total, including 26 feature variables and 1 target variable.

\subsection{Features analysis and elimination}
The participants have provided a total of 700 pieces of information. We utilized the SMOTE (synthetic minority oversampling technique) \cite{gupta2021novel} method to enlarge the size of the dataset, as its size was considerably smaller. The data represents a total of 3286. Then, we figure out the importance of the 26 features by the Chi-squared test \cite{wu2024new} shown in Table \ref{tab2}. Here, this study considers $H0$ to be the null hypothesis and $H1$ denotes the relationship between variables:\\
\\ $H0:$ No relation between the feature and class variable.\\
$H1:$ The target variable and the feature are co-related to each other.\\
\\The null hypothesis is rejected if the p-value revealed by the result is less than a significance threshold (0.05). 
For all characteristics in Table \ref{tab2} with the exception of age, accessing online accounts using multiple devices, and knowledge of cybercrime features, the p-value is less than the significance threshold. So, we reject the null hypothesis for those features where the p-value is less than the significance level. On the other hand, we will not take into consideration factors like age, knowledge of cybercrime, and accessing internet accounts from several devices for further analysis.

\begin{table*}[width=0.58\textwidth]
\centering
\caption{Features importance using Chi-squared test}
\begin{tabular*}{\tblwidth}{l|l}
\hline
\hline
\textbf{Features} & \textbf{P-Value} \\
\hline
\hline
Shared online account access with others & 2.737850e-198 *** \\
Sharing private information on the internet & 2.584012e-185 *** \\
Clicked on spam email links & 5.356233e-183 *** \\
Compulsive buyer & 3.258221e-179 *** \\
Lost money purchasing online commodities & 1.509159e-161 *** \\
Accessed VPN & 7.151082e-160 *** \\
Shared email access & 3.337742e-149 *** \\
Shared private devices & 4.261424e-142 *** \\
Disclose sentiments on social media & 4.729707e-136 *** \\
Weak password & 6.338486e-129 *** \\
Used malware infected flash drive & 2.386920e-114 *** \\
Download unauthorized software & 2.395619e-70 *** \\
Stored credentials on browsers & 3.102031e-68 *** \\
Social media user & 1.848618e-50 *** \\
Maintained privacy on social media & 2.551826e-47 *** \\
Online products purchaser & 3.746547e-41 *** \\
Gender & 1.482072e-25 *** \\
Installed malicious software & 1.645218e-22 *** \\
Permitted ingress in email & 1.769798e-17 *** \\
Receive phishing email & 6.025903e-11 *** \\
Victimized by blackmailing & 9.292576e-07 *** \\
Aware about cybercrime & 1.213865e-06 *** \\
Devices keep updated & 1.071657e-04 *** \\
Age & 3.134269e-01 \\
Accessing online account using several devices & 4.636891e-01 \\
Knowledge about cybercrime & 5.411539e-01 \\
\hline
\end{tabular*}
\label{tab2}
\end{table*}

\begin{figure*}[ht!]
\centering
\includegraphics[width=17.5 cm, height=11.5 cm]{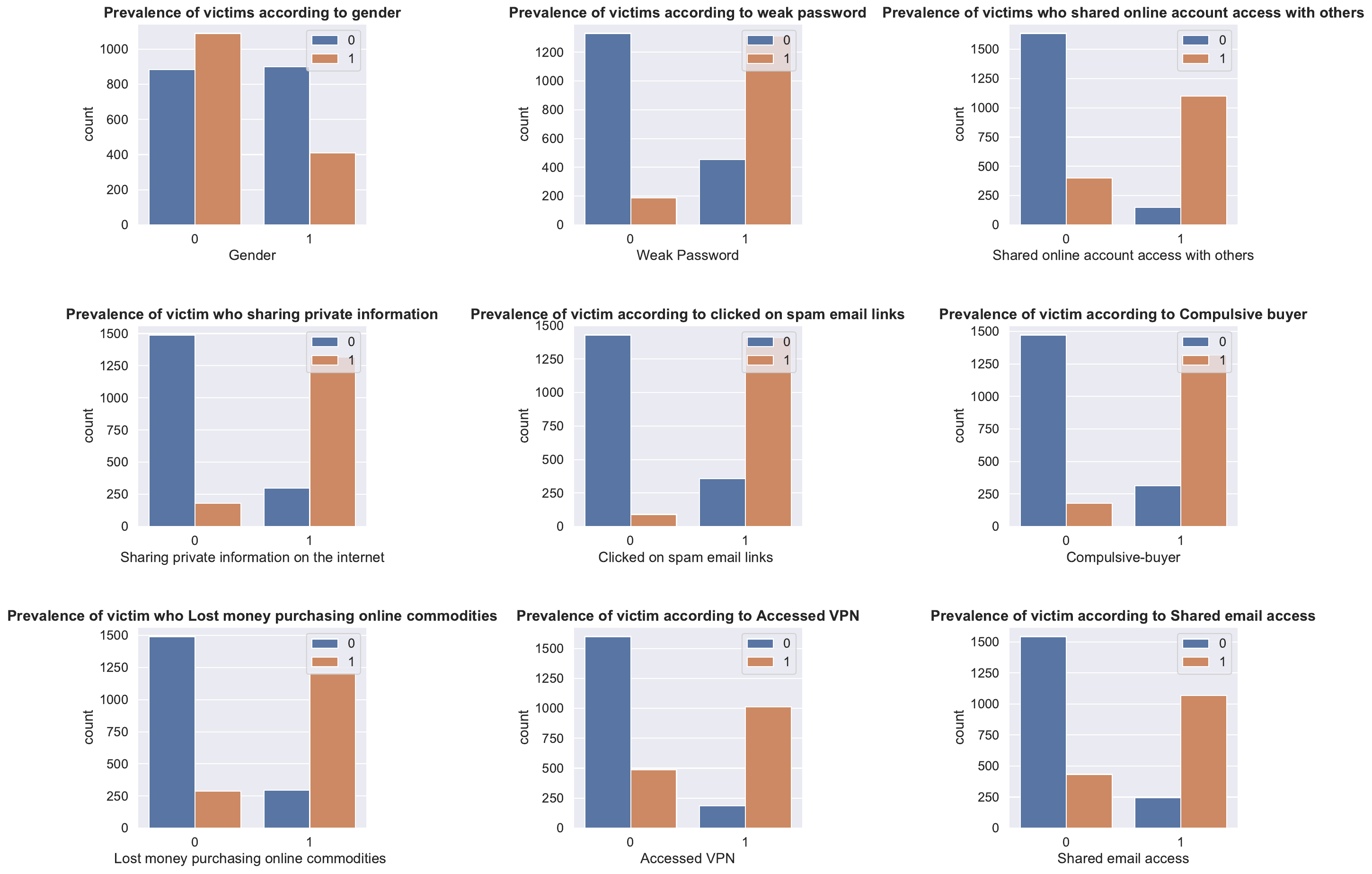}
\caption{The prevalence of the features according to the victim}
\label{fig1}
 \end{figure*}

To eliminate the characteristics with the least impact, Backward elimination \cite{gaugel2024data} of features is performed. Backward elimination is a technique that begins by using all of the features to train and develop the classifier, and then in each step, it provides accuracy and finds the classifier with the best performance. The qualities utilized to create the identified classifier are regarded to be high-impacting characteristics, while the remaining characteristics are eliminated. The approach was carried out by initially training the classifier with all 26 feature variables, monitoring the influence of each variable on the target feature at each step, and deleting the least influential features based on the highest accuracy. The performance of each step has been recorded for comparative purposes but it is difficult to display the performance of the classification ML models trained with each and every feature. Nevertheless, the classification models trained with some features are displayed in Table \ref{tab5}. The predictive classifier trained with 20 features is ultimately selected as the best classifier. The 20 construction-related features were chosen as the most important of the initial 26 features collected during our data collection. 
 
As indicated in Figure \ref{fig1}, we next examined the selected features by Chi-squared test and Backward elimination and also sought to illustrate the prevalence of victims depending on other variables. The subsequent paragraphs explain Figure \ref{fig1}. Female is denoted as 0 and Male as 1 in our dataset. Females are mostly victims of cyber-attacks. On the other hand, there are fewer male victims. The people who use weak passwords have become victims more than the individuals with strong passwords. Next, those who share online account access with others are more victims of cyber attacks than those who do not. Most individuals are aware of sharing their private information on the internet. However, a huge amount of individuals who disclose their personal information on the internet are particularly vulnerable to cyber threats. The people who usually clicked on spam emails or phishing links are more severe victims of cybercrime than those who did not click. we noticed that a huge amount of people are compulsive buyers in which a high proportion of these individuals were victims of cyber assaults. A large number of individuals have lost their money using social media but among those people, the majority have been victims of cyber-attacks. Fewer individuals access VPNs (virtual private networks), nevertheless, among them most people are victims. Eventually, a scanty number of users shared email access, and the greatest number of these users were attacked by cybercrime. 
It is challenging to demonstrate every feature, nevertheless, we tried to display and compare most of the features subject to the goal feature and found that our selected features are hugely impactful according to Figure \ref{fig1}.

\subsection{Training and selection of classifier}
The earliest 26 variables in the dataset are features, whereas variable 27 is the goal variable. The goal variables and feature variables are maintained in distinct data frames. The complete dataset was separated into train datasets comprising 75\%, test datasets comprising 17.55\%, and validation dataset comprising 7.55\%. The train dataset was sent to ML classification algorithms, such as RF, DT, LR, SVM, GB, and GNB\cite{sarker2023machine} to develop the predictive binary classifier. To evaluate the accuracy of classifiers, their output data are compared with the target variable in the test dataset.

As soon as the Backward elimination approach was selected, algorithm training with all 26 characteristics commenced. At each successive step, the models were trained with one fewer feature, and the technique was repeated until the optimal classifier was discovered. Using the Scikit-learn library methods RandomForest Classifier(), DecisionTree Classifier(), Logistic Regression(), SVC(), GradientBoosting Classifier(), and GaussianNB Classifier(), the binary classifiers were trained and produced. After training, these classifiers could predict the outcome of a brand-new data input.

An RF classifier is a well-known ensemble classification technique used in a variety of ML and data science application domains. It is a collection of trees that attempts to compute the outcome by generating several DTs from randomly selected data samples. Voting picks the answer to the tree with the highest degree of precision. Large, randomly distributed datasets are analyzed with high precision. In our case, algorithms are trained to utilize the properties of the data collection. The algorithm then predicts the class based on the training data and whether the instance fits victim or non-victim characteristics based on a vote. A DT classifier is a tree-structured classifier and constructs a classification model in which internal nodes stand in for the attributes of a dataset,  branches for the decision-making processes, and leaf nodes for the results. It divides a dataset into progressively smaller subgroups while simultaneously developing an associated DT. The DT uses the attributes from our dataset as input, and the leaf nodes forecast whether the person falls into the victim or non-victim group.
\begin{figure*}[ht!]
\centering
\includegraphics[width=17.5 cm, height=11.5 cm]{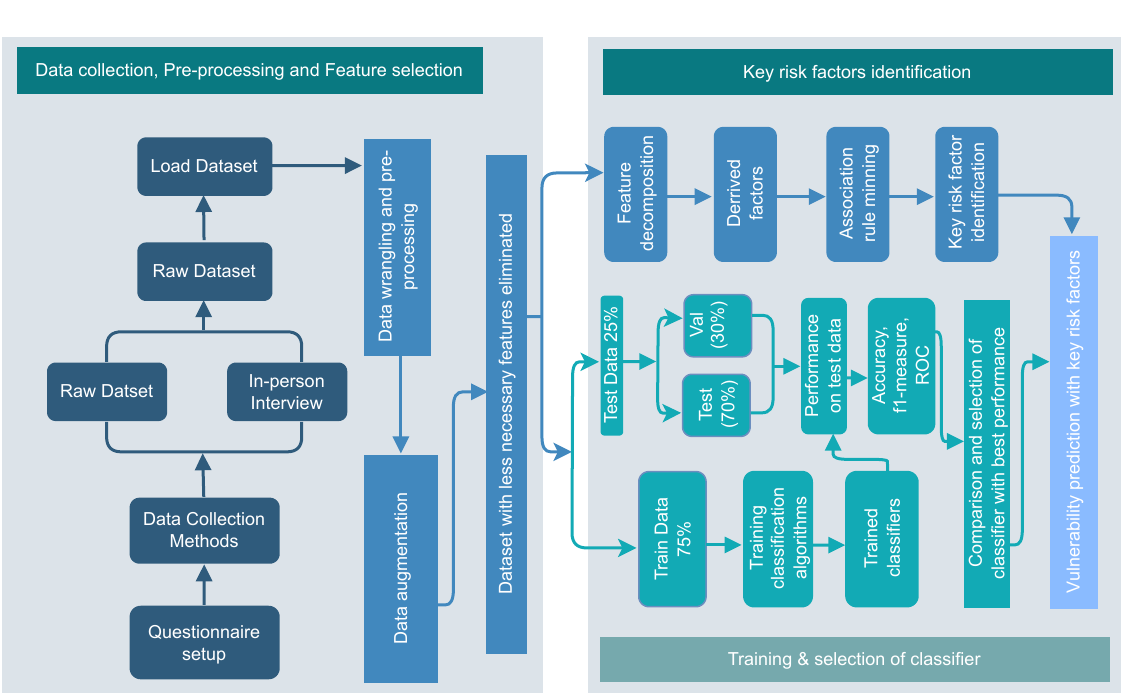}
\caption{Overview of the proposed framework.}
\label{flowfig}
 \end{figure*}

LR is the prevalent probabilistic-based statistical model used to handle classification problems in ML. It commonly employs a logistic function, often known as the theoretically defined sigmoid function, to estimate the probability. It works well when the data can be split linearly to predict the output. Output values are binary values that may also be considered categorization classes. In the training phase with our training data, the algorithm selects a regression line based on our input characteristics. When an individual's data is supplied, the model determines the location of the feature based on the learned regression line. The activation function then determines whether the instance corresponds to the victim category or not.  
The Support Vector Classifier (SVC) technique generates a hyperplane to divide n-dimensional space into classes, allowing us to conveniently place new data points in the appropriate category in the future. In our dataset, the algorithm determines whether the input attributes belong to the victim or non-victim part of the training hyperplane and guesses the class appropriately.
GB classifiers are a type of ML algorithm that integrates a large number of poor learning models to produce a robust prediction model. As the name indicates, gradient-boosting classifiers are methods that are utilized for classification applications. In our dataset, we trained our input dataset using this classifier technique and built a model to predict if a person is a cyber-attack victim or not.
GNB classifiers are an ensemble learning approach and a collection of ML algorithms that integrate several weak algorithms to construct a robust prediction model; it works with loss function, weak learner, and adaptive model. The GNB classifier may be utilized if the target characteristic is binary. The performance of the GNB classifier will be enhanced by reducing the over-fitting issue and performing regularization. Considering our situation, it determines the likelihood associated with the input attributes based on the characteristics of the individuals. Finally, it forecasts the output for these characteristics based on a greater likelihood.
The accuracy score of various classification methods was recorded with the precision, recall, and weighted f1-measure, and the ROC curves were also computed as shown in subsection \ref{ROC_AUC}. 

\subsection{Identifying key risk factors}
\label{risk_factor}
After completing feature selection, eradication, and determining the greatest classifier, 20 characteristics were utilized to train the classifier with the highest performance.  If we could build a mechanism to determine which of these 20 features has the most
\begin{table*}[width=0.79\textwidth]
\caption{Dissolution of essential features into factors}
\centering
\begin{tabular*}{\tblwidth}{l|C|l}
\hline
\hline
{\bfseries Features} &   {\bfseries Factor} & {\bfseries Factors}\\
 &   {\bfseries no} & \\
\hline
\hline
{\itshape Weak password } & 1  & utilized weak passwords such as 1,2,3,4,5,6.. on local \\
                         &      & machines or internet accounts.\\
                         & 2  & never used passwords such as 1,2,3,4,5,6 on \\
                         &      & local machines or internet accounts.\\

\hline

{\itshape Victimized by } & 3  & poorly influenced by blackmailing\\
{\itshape blackmailing} & 4  & never victimized by blackmailing.\\
\hline
{\itshape Lost money purchasing} & 5  & misplaced money by purchasing online \\
& & commodities\\
{\itshape online commodities}  & 6  & never misplaced money by purchasing\\
& & online commodities\\
\hline
{\itshape Compulsive buyer} & 7  & have sufficient control over online product purchases. \\
                         & 8  & compulsively purchase the online product.\\
\hline
{\itshape Disclose sentiments on} & 9 & Always disclose sentiments in public places \\
 {\itshape social media}                & &like social media. \\
    & 10 & never shared sentiments on social media.\\

\hline               
{\itshape Download unauthorized}  & 11 & Always download software from third-party sources.\\
{\itshape  software} 
                & 12 & never downloaded software from third-party sources.\\
\hline
{\itshape Shared email access} & 13 & shared personal information like email ID access \\
& &  with others. \\ 
& 14 & never shared email ID access with others. \\
\hline
{\itshape Permitted ingress in} & 15 & permitted another user access to the email account.\\
{\itshape email} 
& 16 & never permitted another user access to the email \\
& & account. \\
\hline
{\itshape Social media user} & 17 &  uses social media like Facebook, Twitter, etc.\\
& 18 & have no social media account.\\
\hline
{\itshape Clicked on spam } & 19 & clicked on spam or fraudulent links in emails.\\
{\itshape email links} & 20 & never clicked on spam or fraudulent links in emails.\\
\hline

{\itshape Shared private devices} & 21 & Shared private devices with anyone else.\\

& 22 & never Shared private devices with anyone else.\\
\hline
{\itshape Accessed VPN } & 23 & accessed VPN persistently\\
& 24 & does not access VPN persistently.\\
\hline
{\itshape Sharing private } & 25 & share confidential information on social media, like \\
{\itshape information on } & & images or bank account numbers.\\

{\itshape the internet} & 26 & never shared confidential information on social\\
& &  media.\\
\hline
{\itshape Installed malicious } & 27 & installed malicious software on the devices.\\
 {\itshape software}
& 28 & never installed malicious software on the devices.\\
\hline
{\itshape Shared internet account} & 29 &  shared internet account access with others.\\
{\itshape  access  }
& 30 & never shared internet account access with others.\\
\hline
{\itshape Used malware infected } & 31 & used malware-infected flash drives in \\
{\itshape flash drive} & &  personal devices.\\
& 32 & never use malware-infected flash drives \\
& & on personal devices.\\
\hline
{\itshape Receive phishing email } & 33 & received always phishing emails\\
& 34 & never received any phishing emails\\
\hline
{\itshape Online products purchaser } & 35 & online products purchaser \\
& 36 & not an online products purchaser.\\
\hline
{\itshape Aware about cybercrime} & 37 & aware of cybercrime.\\
& 38 & not aware of cybercrime.\\
\hline
\end{tabular*}
\label{tab3}\\
\end{table*}
influence on cyber risks and how they impact cyber risks, it would be valuable for comprehending the cyber risk scenario as a whole and taking action. This issue was resolved by separating these 20 characteristics into unique elements. As each of these characteristics is categorical, it may be broken down into several factors. Then, we identified the most influential attributes via association rule mining.

For example, the "Misplaced money by purchasing online commodities" feature was generated from item Q13 in subsection \ref{subsec:3.1} , Q13: Has the user misplaced money by purchasing or vending commodities online? This question has two alternative responses in our questionnaire:  1. Has misplaced money by purchasing or vending commodities online, 2. Has not misplaced money by purchasing or vending commodities online. Because the "Lost money by purchasing online commodities" feature might be classified under any of these two categories, we refer to them as the feature's causes. Thus, the characteristic "Lost money by purchasing online commodities" was split into these two factors. Similar to this feature, each of these 20 features is subdivided into the number of categories they represent. All of these elements are detailed in Table \ref{tab3}, where all 38 dissolved factors are displayed.

Now, to determine the most influential cyber risk variables, we assess these 38 elements. Association rule mining\cite{sarker2024ai} is utilized to determine the 38 most significant factors listed in subsection \ref{4.5} for analysis. By employing association rule mining, this study can determine which of these 38 factors are simultaneously present in the cases of victims. Association rule mining produces rules based on the relationships among the factors. Each rule is assigned a confidence rating that represents its probabilistic strength. Apriori is an association rule mining algorithm that finds simultaneous items in a dataset by computing their frequency\cite{sarker2023multi}. This algorithm was utilized to determine which characteristics were simultaneously present in victim responders. Various factors were extracted from mining rules, and through using support, lift, and confidence between the factors of any rule, their performance has been evaluated. The estimated relationship between left- and right-hand side characteristics is designated as a rule. 
In a rule: if 'A', then 'B', confidence = 80\%, where 'A' is the antecedent and 'B' is the consequent, the probability of B being true in the instance of an individual when A is true is 80\%.
Here, 'A' might be an individual or a combination of many factors from the 38 derived, and 'B' indicates when the respondent is the victim. Using the algorithm, rules were developed and evaluated based on the metrics stated in subsection \ref{output_generation}; the selected rules are presented in subsection \ref{4.5}.

\subsection{Model parameters}
Within this section, this study presents a summary of the parameters utilized in various modeling and experimental approaches. In each phase of classifier comparison, algorithms were incrementally trained using 75 percent of the dataset with 1 to 26 feature variables, and the single training variable and the remaining 25 percent of the dataset were employed for testing and validating. The comparison was beneficial for identifying a credible predictor. 
\begin{table*}[width=0.73\textwidth]
\caption{Parameters employed in several procedures}
\centering
\begin{tabular*}{\tblwidth}{l|l}
\hline
\hline
{\bfseries Algorithms} &   {\bfseries Parameters}\\
\hline
\hline
{\itshape RF }  & n\_estimators = 10, random\_state = 42 , criterion ``gini" \\
{\itshape DT } & criterion = ``gini", splitter=``best", random\_state=22 \\
{\itshape LR } & penalty= ``l2",  dual= False, random\_state=22 \\
{\itshape SVC } & kernel=``poly", gamma=``scale"\\
{\itshape Gradient Boosting} & criterion =``friedman\_mse", learning\_rate=0.1, n\_estimators=100 \\
{\itshape Gaussian NB } & priors = None, var\_smoothing=1e-9 \\
{\itshape Apriori } & min support=0.25, min confidence=0.8, numRules=10000 \\
\hline
\end{tabular*}
\label{tab4}
\end{table*}
\\

Using the Apriori method, significant risk variables have been uncovered. The 20 crucial characteristics have been refactored into 38 factors that have been submitted to the Apriori association rule mining method. From the extracted rules, the rules with the highest level of confidence were determined. Table \ref{tab4} displays the significant parameters provided by the above-mentioned algorithms.

\subsection{Output generation} \label{output_generation}
In this study, to discover the major factors in the classifiers, the Chi-square test and Backward elimination strategy are employed to select important features and to eliminate less significant features, and then the remaining features are dissolved and analyzed. We measured accuracy using the Scikit-learn function accuracy score(), which uses the predicted output of the classifiers in the test dataset. As test data is one-fourth of the original dataset, we split the test dataset into 70\% and 30\% and preserved the remaining 30\% dataset for validation. The accuracy score is a useful indicator for determining the effectiveness of the classifier on data from a new individual. Using the method of Backward elimination, we compared the accuracy scores of all 6 algorithms with characteristics to choose the classifier with the highest accuracy. Beginning with all 27 feature variables and reducing them to a single variable, we concluded that the RF classifier trained with 20 features had the maximum accuracy.
As it is a binary classification strategy, various measures such as accuracy, recall or TPR (True Positive Rate), TNR (True Negative Rate), and f1-measure and their implications are required to validate the model's effectiveness. To comprehend these terminologies, a confusion matrix must be elucidated. In this study, a confusion matrix is a 2x2 table in which each column and row represents a unique occurrence of the predicted class and actual class,  appropriately. Figure \ref{fig3} displays a confusion matrix, and its expressions are defined below:

\begin{figure}[ht!]
\centering
\includegraphics[width=6.5 cm, height=4.5 cm]{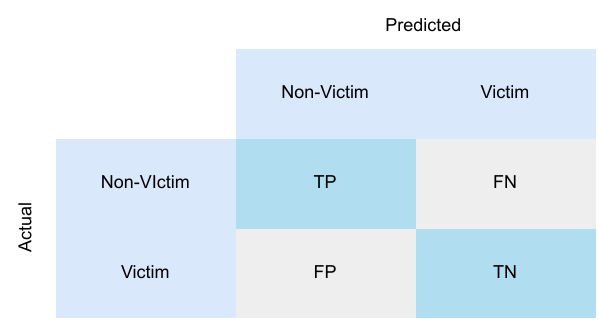}
\caption{Confusion Matrix based on Victim and Non-Victim}
\label{fig3}
 \end{figure}

\paragraph{TP (True Positive):} How many of the events in the dataset are accurately predicted as non-victims by the classifier.
\paragraph{FP (False positive):} How many outcomes are wrongly predicted as non-victims by the classifier when they are victims in the dataset.
\paragraph{FN (False negative):} How many outcomes are incorrectly predicted as victims when they are non-victims in the dataset.
\paragraph{TN (True negative):} How many outcomes in the dataset are accurately predicted by the classifier to be victims.

\begin{equation}
Precision = \frac{TP}{TP + FP}
\end{equation}

\begin{equation}
Recall = \frac{TP}{TP + FN}
\end{equation}

\begin{equation}
TNR = \frac{TN}{TN + FP}
\end{equation}

\begin{equation}
F1 = \frac{2 \cdot Precision \cdot Recall}{Precision + Recall}
\end{equation}
\\
By assessing accuracy, recall, TNR, and the f1-score, it is possible to determine how well the classifier predicts victim and non-victim classifications. We determined these values, drew the ROC (receiver operator characteristic) curve, and observed the AUC (area under the curve) value. ROC curves are widely used to graphically illustrate the relationship or trade-off between clinical sensitivity and specificity for every feasible cut-off for a test or group of tests. False Positive Rate (1-TNR) is plotted along the X-axis, while True Positive Rate (Recall) is plotted along the Y-axis. AUC is the two-dimensional area under the ROC curve used to compare the performance of a prediction model. A test's area under the curve (AUC) can be used to measure its discriminating capacity, i.e., how well the test performs in a certain clinical context. In general, a larger AUC value denotes enhanced predictive ability. By comparing all of these metrics, the most efficient classifier is selected.
As stated in subsection \ref{subsection4.1}, the classifier with the best performance utilized 20 characteristics. To identify key factors, the Apriori algorithm was utilized to construct association rules based on these characteristics. This algorithm determines the relation between characteristics in terms of three metrics: support which determines the frequency of any component or factor combination within the dataset, the lift which is the percentage of antecedents and consequences that are present together in the dataset instead of being present alone, and confidence that indicates the potential presence of a consequence when the antecedent is already present. Rules developed based on these factors were assessed based on these metrics for comprehending the relevance of the rules statistically. The rules that reached the threshold values based on these metrics were evaluated, while the others were eliminated. For identifying important risk factors, this research analyzed rules with greater than 80\% confidence.

\section{Experimental results} \label{sec:4}
The first step is the reduction of characteristics that are less vital in cyber risk. Using the Chi-squared test and  Backward elimination method, 20 essential features were identified for training the best-performing classifier, and the rest were excluded from the dataset. Accuracy scores, ROC curve analysis, precision, recall, and f1 measures are applied for classifier selection. By dividing the attributes used in this classifier into elements, association rules were constructed from the factors necessary to identify the most essential traits.

\subsection{Accuracy scores} \label{subsection4.1}
\begin{figure}[ht!]
\centering
\includegraphics[width=7.5 cm, height=6.5 cm]{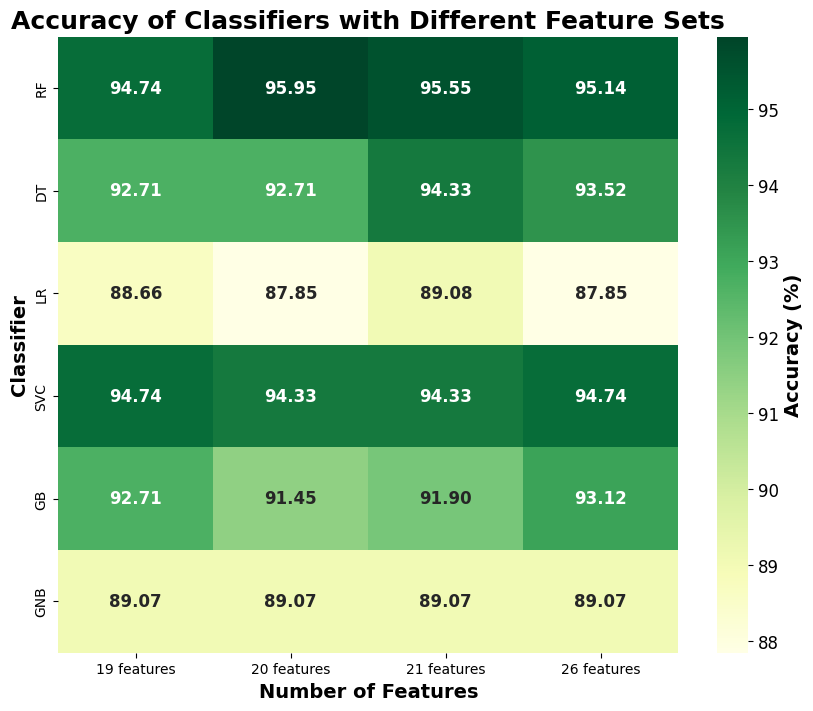}
\caption{Heatmap of the  classifier's accuracy}
\label{fig55}
 \end{figure}

As feature selection was accomplished using the Chi-squared test and the Backward elimination approach, each step's performance was recorded so that classifiers could be compared and analyzed. The classifier with the best performance in each iteration was found and then compared to other models with the best performance in other iterations. In Table \ref{tab5}, the classification accuracy ratings are given in 4 steps. Figure \ref{fig55} represents the accuracy heatmap.

\begin{table*}[width=0.69\textwidth]
\centering
 \caption{Percentage(\%) accuracy of classifiers trained with features picked by Chi-squared test and Backward elimination}
\begin{tabular*}{\tblwidth}{C C C C C }
\hline
 \hline
{\bfseries Classifier trained with \ \ } & {\bfseries 19 features  } & {\bfseries 20 features  } & {\bfseries 21 features  } & {\bfseries 26 features }\\
\hline
\hline
RF &	94.74 &	95.95 &	 95.55 &	95.14\\
DT &	92.71 &	92.71 &	94.33 &	93.52 \\
LR &	88.66 &	87.85 &	89.08 &	87.85\\
SVC	& 94.74 &	94.33 &	94.33 &	94.74\\
GB &	92.71 &	91.45 &	91.90 &	93.12\\
GNB	& 89.07 &	89.07 &	89.07 &	89.07\\
\hline
\end{tabular*}
\label{tab5}
\end{table*}

Displaying the outcomes for all 26 stages is not only impractical but also expensive. This study has given the accuracy scores of the models trained with 19 20, 21, and lastly, all 26 features to demonstrate that the performance of the classifier trained with 20 feature variables is superior to the others. Comparing the best performance across all columns, the RF classifier trained with 20 feature variables had the greatest accuracy score of 95.95 percent. As discussed previously, this accuracy score indicates how well a model is on a given dataset. As the size of the dataset is (n= 3286) and data is divided into the train, validation, and test datasets, this accuracy score indicates how well the model can predict the class of 247 individuals within the validation dataset.
\subsection{ROC curve and AUC value analysis} \label{ROC_AUC}
The RF classifier has the highest performance among classifiers trained with the most important 20 features selected by the Chi-squared test, as shown by analyzing the data in Table \ref{tab5} and Figure \ref{fig55}. In addition to the classification accuracy, it must compare the ROC curves for each of these classifiers to comprehend their performance in terms of sensitivity and specificity. AUC values were utilized to compare classifiers using ROC curves, which offer a good representation of the discriminating capacity.AUC values are necessary in this study because it needed to understand the performance of the classifiers in the case of both victim and non-victim classes, where the accuracy score may not describe the complete situation given that the dataset of this study contains a greater number of responses from non-victims. Figure \ref{fig4} displays the ROC curves and AUC values for classifiers listed in Table \ref{tab6}.
The ROC curve for each case is depicted in Figure \ref{fig4}, representing classifiers trained with 19, 20, 21, and all 26 features, respectively. The AUC values for each classifier are shown in the bottom right corner of each graph. By evaluating AUC values in conjunction with accuracy scores, it may be possible to reach a persuasive conclusion in selecting the ideal model for this study. Analyzing the AUC values of classifiers with the highest accuracy score in each case highlighted in Table \ref{tab6} and Figure \ref{fig4}, the RF classifier has a great combination of classifiers trained with 19 features with an accuracy score of 94.74\% and an AUC value of 0.99. Again, RF has the best combination for 21 characteristics, with an accuracy score of 95.55\% and an AUC value of 0.98. RF has the greatest combination of 20 variables, with an accuracy of 95.95\% and an AUC value of 0.98. Eventually, when trained with all 26 features, the RF classifier has good accuracy with a score of 95.14\% and the same AUC value as before (0.99).

\subsection{Precision and Recall value analysis}
A RF classifier trained with 20 feature variables offers the greatest combination of accuracy and AUC among all classifiers. Table \ref{tab6} describes each of the above-mentioned classifiers for each of the binary classifications victim and non-victim in terms of additional metrics, including precision and recall. Precision describes the chance of correctly classifying a positive class, whereas recall describes the model's sensitivity toward recognizing the positive class. The F1-Score is the weighted average of precision and recall. Thus, this score takes into consideration both false positives and false negatives. These scores are necessary in the situation of an imbalanced dataset in which the items inside the dataset are not distributed equally among the classes, in our case binary classes.
\begin{figure*}[ht!]
\centering
{\includegraphics[width=6cm]{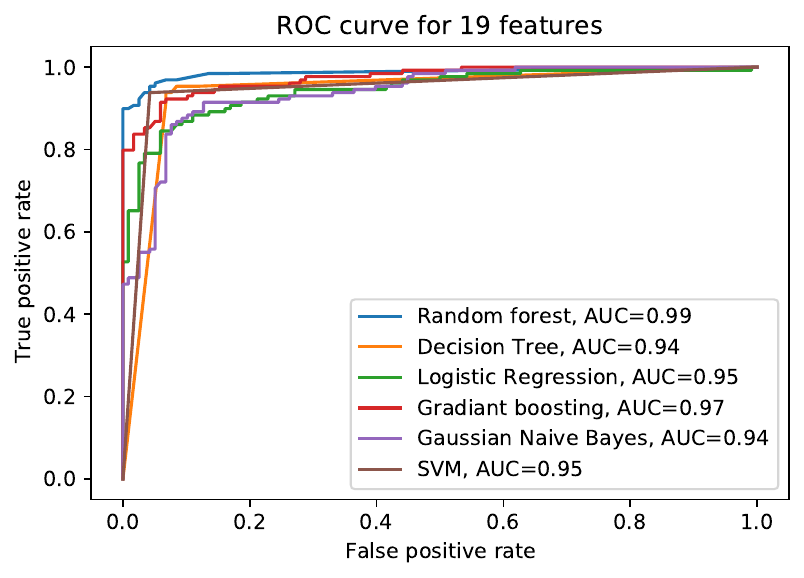}}
\quad 
{\includegraphics[width=6cm]{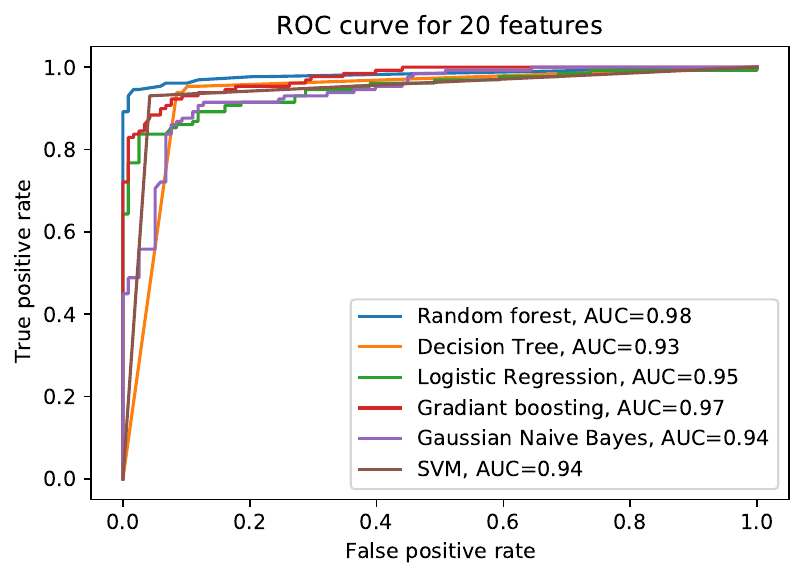}}

{\includegraphics[width=6cm]{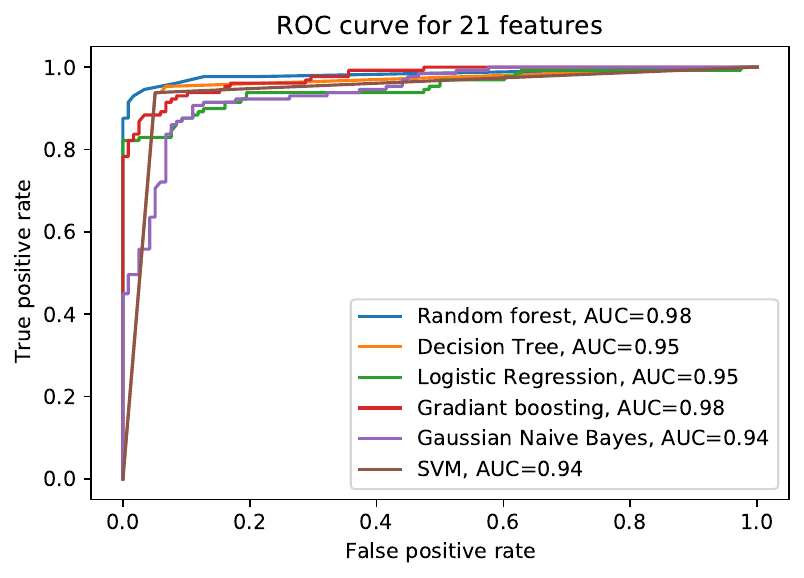}}
\quad 
{\includegraphics[width=6cm]{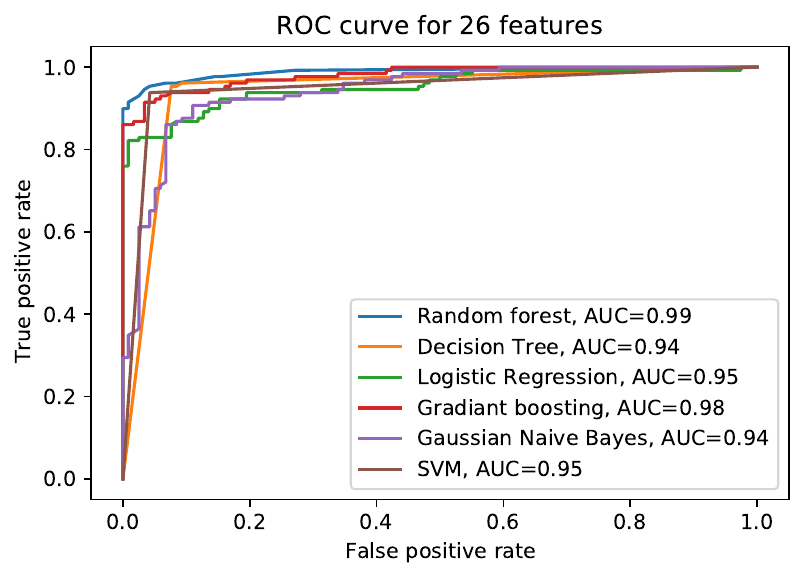}}

\caption{ROC curves for classifiers trained using the 19, 20, 21 features, and all 26 feature variables}
\label{fig4}
\end{figure*}
In Table \ref{tab6}, RF represents Random Forest, DT indicates the Decision Tree, LR denotes Logistic Regression classifiers, SVC stands for Support Vector Classifier, GB denotes Gradient Boosting classifier, and GNB represents Gaussian Naive Bayes. Precision and recall in Table \ref{tab5} for classifiers are listed in Table \ref{tab6} and Figure \ref{fig4} respectively. The weighted average is required since the dataset is somewhat imbalanced. RF and SVC have the highest weighted average f1 score of 0.95 among classifiers trained with 19 feature variables; with 20 features, RF has the highest f1 score of 0.96; with 21 feature variables, RF has the best weighted average f1 score of 0.96; and with 26 features, RF and SVC have the highest f1-score of 0.95.



\begin{table*}[width=0.79\textwidth]
\centering
\caption{Precision and Recall values for applied classifiers}
\begin{tabular*}{\tblwidth}{C C C C C C C C C C}
    \hline
    \hline
    {\bfseries Classifier } &  & \multicolumn{2}{c}{\bfseries 19 Features \ } & \multicolumn{2}{c}{\bfseries 20 Features \ } & \multicolumn{2}{c}{\bfseries 21 Features \ } & \multicolumn{2}{c}{\bfseries All Features } \\
    \hline
    \hline
    & {\itshape Class } & { \itshape Prec } & {\itshape Rec} &  { \itshape Prec } & {\itshape Rec} &  { \itshape Prec } & {\itshape Rec} &  { \ \itshape Prec \ \ } & {\itshape Rec}\\
    \hline
    \itshape RF & { Non-Victim } & 0.93 & 0.96 & 0.94 & 0.97 & 0.93 & 0.97 & 0.92 & 0.98\\
     & Victim & 0.96 & 0.94 & 0.98 & 0.95 & 0.98 & 0.94 & 0.98 & 0.92\\
    \hline
    \itshape DT & { Non-Victim } & 0.95 & 0.90 & 0.95 & 0.90 & 0.95 & 0.93 & 0.95 & 0.92\\
     & Victim & 0.91 & 0.95 & 0.91 & 0.95 & 0.94 & 0.95 & 0.92 & 0.95\\
    \hline
    \itshape LR & { Non-Victim } & 0.88 & 0.89 & 0.87 & 0.88 & 0.87 & 0.91 & 0.87 & 0.88\\
     & Victim & 0.90 & 0.88 & 0.89 & 0.88 & 0.91 & 0.88 & 0.89 & 0.88\\
    \hline
    \itshape SVC & { Non-Victim } & 0.93 & 0.96 & 0.93 & 0.96 & 0.93 & 0.95 & 0.93 & 0.96\\
     & Victim & 0.96 & 0.94 & 0.96 & 0.93 & 0.95 & 0.94 & 0.96 & 0.94\\
    \hline
    \itshape GB & { Non-Victim } & 0.91 & 0.94 & 0.89 & 0.93 & 0.90 & 0.93 & 0.91 & 0.95\\
     & Victim & 0.94 & 0.91 & 0.94 & 0.90 & 0.94 & 0.91 & 0.95 & 0.91\\
     \hline
     \itshape GNB & { Non-Victim } & 0.86 & 0.92 & 0.86 & 0.92 & 0.86 & 0.92 & 0.86 & 0.92\\
     & Victim & 0.92 & 0.87 & 0.92 & 0.87 & 0.92 & 0.87 & 0.92 & 0.87\\
     \hline
  \end{tabular*}
  \label{tab6}
\end{table*}

Comparing all classifiers, it can be shown that the RF classifier trained with 20 feature variables has the most accurate ability to distinguish between the victim and non-victim classes, with an accuracy score of 95.95\%, a weighted f1 average of 0.96, and an AUC value of 0.98. Analyzing the results, an accuracy score of 95.95\%, which represents the percentage of properly classified data by the classifier, may be deemed an excellent score. Both the precision and recall scores for the victim and non-victim groups were considerably above 0.85, indicating that the classifier accurately predicts for both classes.
\subsection{Error analysis}
\begin{figure}[ht!]
\centering
\includegraphics[width=6.5 cm, height=5.5 cm]{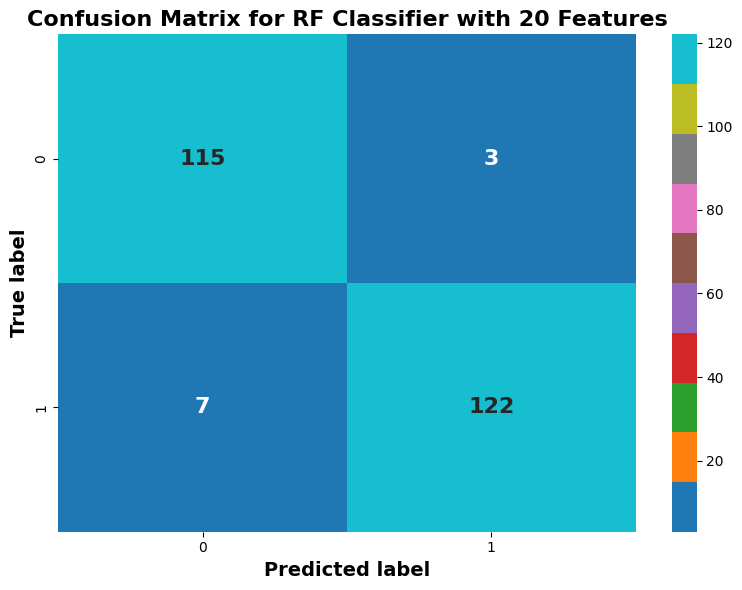}
\caption{The analysis of features}
\label{fig6}
 \end{figure}

It would be easier to comprehend the performance of the chosen classifier if the error analysis was included. Error analysis is simply explicable using a confusion matrix. Figure \ref{fig6}  is the confusion matrix for the RF classifier trained with 20 features.

From Figure \ref{fig6}, we can see that, out of a total of 247 predictions, 115 respondents were accurately recognized as non-victims, while just 3 non-victim responders were wrongly labelled as victims. In the instance of victim predictions, 7 victims are wrongly labelled as non-victims, while the remaining 122 victims are accurately recognized.

\subsection{Key risk factors} \label{4.5}
The 20 variables utilized for training classification algorithms to construct the appropriate classifier will be examined to determine their influence and identify key factors. In the subsection \ref{risk_factor}, the procedure for dissolving 20 features into 38 factors is outlined. Examine the factors listed in Table \ref{tab3} to determine which have a significant influence on cyber risk. We accomplished this by mining association rules between these 38 factors and examining the rules about "victim" persons among the urban youth demographic in Bangladesh. Important rules were grouped based on confidence levels exceeding thresholds, as seen in section \ref{output_generation}. Table \ref{rule} has a comprehensive study of these rules.

\begin{table*}[width=0.79\textwidth]
\caption{Rules derived from the analysis of the 38 factors presented in table 3}
\centering
\begin{tabular*}{\tblwidth}{C|C|C} 
\hline
\hline
{\bfseries Class} & {\bfseries Rules} & {\bfseries Confidence} \\ [0.1ex]
\hline
\hline
\multirow {20}{*}{\textit{Victim}} & If (compulsively purchase the online product, download software from & 82.0\%  \\
 & third-party sources, installed malicious software on the devices) then  &  \\ [2ex]

& If (utilized weak passwords on local machines or internet accounts, &  85.0\% \\ 
& download software from third-party sources, sharing private information &\\&
on the internet, installed malicious software on the devices) then  & \\[2ex]
 
& If (utilized weak passwords on local machines or internet accounts, & 98.0\% \\ 
& misplaced money by purchasing online commodities, compulsively  &  \\
& purchasing the online product, downloading software from third-party &\\& 
sources, sharing private information on the internet, using &\\&
malware-infected flash drives)then  &  \\ [2ex]

& If (buying online products, compulsively purchase the online product, &  88.0\% \\ 
& download software from third-party sources, sharing private information &\\&
on the internet) then  &  \\ [2ex]
& If (compulsively purchase the online product, clicked on spam or &  100.0\% \\ 
& fraudulent links in emails, accessed VPN persistently, used &\\& 
malware-infected flash drives)then &  \\ [2ex]

& If (utilized weak passwords on local machines or internet accounts, & 90.0\% \\
& buying online products, compulsively purchasing the online product,
&\\& downloading software from third-party sources, sharing private &\\&
information on the internet)then & \\ [2ex]

& If (permitted another user access to the email account, click on & 93.0\% \\ 
& spam or fraudulent links in emails, sharing private information on &\\& 
the internet, shared internet account access with others, used &\\& 
malware-infected flash drives)then  &  \\ [2ex]

& If (buying online products, social media user, clicked on spam or  & 83.0\% \\ 
& fraudulent links in emails) then &  \\ [2ex]
& If (buying online products, misplaced money by purchasing online & 98.0\% \\ 
&  commodities, compulsively purchasing online products, clicking on  &\\&
spam or fraudulent links in emails, downloading software from &  \\
& third-party sources, sharing private information on the internet, &\\&
using malware-infected flash drives) then  & \\ [2ex]

& If (utilized weak passwords on local machines or internet accounts, &\\&
misplaced money by purchasing online commodities, clicking on & 96.0\% \\ 
&  spam or fraudulent links in emails, downloading software from &  \\
& third-party sources, sharing private information on the internet,&\\&
 shared internet account access with others, used malware-infected &\\& 
 flash drives)then  & \\ [2ex]

 \hline
\end{tabular*}
\label{rule}
\end{table*}
These ten essential rules, taken from the pool of all rules mined from the 38 variables, will provide us with a full grasp of the most significant risk factors. The components on the left are referred to as antecedents, while those on the right are alias consequences. In Table \ref{rule}, the first rule indicates that an individual has an 82.0\% chance of being a victim if they compulsively purchase an online product, download software from third-party sources, and install malicious software on the devices. Then, respondents who utilized weak passwords on local machines or internet accounts, downloaded software from third-party sources, shared private information on the internet, and installed malicious software on the devices have confidence with 85.0\% being victims. The 98.0\% confidence was seen among respondents who used passwords such as 1,2,3,4,5,6, etc. for any online account, misplaced money by purchasing online commodities, compulsively purchased online products, downloaded software from third-party sources, shared private information on the internet, used malware-infected flash drives. In the case of responders who buy online products, compulsively purchase the online product, download software from third-party sources, sharing private information on the internet, confidence is 88.0\% of being a victim. For those who compulsively purchase online products, click on spam or fraudulent links in emails, access VPN persistently, and use malware-infected flash drives, victim confidence increases to 100.0\%. If a person utilized weak passwords on local machines or internet accounts, bought online products, compulsively purchased the online product, downloaded software from third-party sources, and shared private information on the internet, has a 90.0\% chance of being a victim. If a respondent permitted another user access to the email account, clicked on spam or fraudulent links in emails, shared private information on the internet, shared internet account access with others, and used malware-infected flash drives, he or she has a 93.0\% chance of being a victim. If a respondent buying online products, or a social media user, clicked on spam or fraudulent links in emails, they have an 83.0\% chance of being a victim. The respondent has a 98.0\% confidence of being a victim if he or she buys online products, misplaced money by purchasing online commodities, compulsively purchases online products, clicks on spam or fraudulent links in emails, downloads software from third-party sources, shares private information on the internet, using malware-infected flash drives. If an individual used weak passwords on local machines or internet accounts, misplaced money by purchasing online commodities, clicked on spam or fraudulent links in emails, downloaded software from third-party sources, shared private information on the internet, shared internet account access with others, and used malware-infected flash drives, have a 96.0\% confidence of being the victim.
We identify the main risk factors for cyber attacks in Bangladesh by calculating the frequency of elements appearing in crucial rules. The factors are social media users, compulsive buying, clicking on spam links, sharing personal data on the internet, using malware-infected pen drives, misplaced money by purchasing online commodities, downloading unauthorized software, sharing internet account access with others, buying online products, permitted another user access to the email account, accessed VPN persistently, installed malicious software on the devices and using weak passwords. Thus, our proposed approach determines the predictive classifier with the greatest performance after thorough most important feature analysis and also identifies crucial elements among the characteristics employed in the buildup of that classifier.
\section{Discussion} \label{sec:5}
The socioeconomic elements discovered in this study highlight the intricate relationship between societal characteristics and cyber security concerns. These elements, which range from individual behaviours to social standards, substantially impact cyber threat vulnerability and impact. Understanding these aspects offers a more nuanced perspective on cyber risk, emphasising the importance of a holistic cybersecurity strategy that extends beyond technological solutions. It emphasises the necessity of considering socioeconomic factors when developing and executing cyber security policies, ensuring that interventions are targeted to the individual vulnerabilities and needs of varied communities. The findings of this investigation highlight the potential of the ML-based model for predicting cyber security risk and demonstrate the key risk factors of individuals’ cyber risk. The evaluation of the proposed model on real-world datasets demonstrates its effectiveness in accurately identifying the primary risk factors associated with cyber threats. Our proposed model Random Forest algorithm achieved the highest accuracy with 20 features, 95.95\%. This study also showed the relationship among selected features using the Apriori algorithm with Confidence (above 80\%) according to the Victim. The significance of various socioeconomic factors was measured using a questionnaire that was developed specifically for this study, and the results showed that these factors can play a critical role in predicting cybersecurity risk.

The use of popular ML classification algorithms such as DT, RF, SVM, LR, GB Classifier, and GNB allowed us to construct predictive models that can accurately identify the primary risk factors of cyber threats. The extraction of association rules using the popular Apriori algorithm from important attributes selected by the Chi-squared test and elements further confirms the potential of this framework to predict cybersecurity risk.
This study provides valuable insights for organizations and individuals to mitigate the risk of cyber attacks. The proposed framework and significant association rules can be used as a tool for organizations to identify the primary risk factors associated with cyber security and take appropriate measures to mitigate the risk. The framework can also be useful for individuals to assess their risk and take necessary precautions to secure their digital assets.
\section{Conclusion and future work}\label{sec:6}

This work has highlighted the importance of considering socioeconomic factors in predicting cyber security risk through a data-driven analysis. The framework is created by gathering actual data to verify its effectiveness in a real-world context. This study has derived significant association rules between the risk causes to pinpoint the main risk elements of particular cyber security. This proposed framework serves as a valuable tool for organizations and individuals to assess and manage the risk of cyber attacks, and it has the potential to contribute to the development of more effective methods for predicting and managing cyber threats. The overall primary purpose of this study is to identify a person's cyber security risk to repel attacks before they become serious. Future research can be focused on further refining and improving the proposed framework, as well as exploring other potential risk factors that could be used to predict cybersecurity risk. 

\section*{Declaration:} {The authors declare that they have no conflict of interest.}

\bibliographystyle{model1-num-names}
\bibliography{cas-refs}

\end{document}